\documentclass[12pt,letter]{article}

\usepackage{graphicx, epsfig, color}
\def\delew{\Delta_{EW}}
\def\msusy{m_{SUSY}}

\def\to{\rightarrow}

\def\bi{\begin{itemize}}
\def\ei{\end{itemize}}
\def\te{\tilde e}

\def\tu{\tilde u}
\def\sps1ap{SPS1a$^\prime$}
\def\c1p{C1$^\prime$}

\def\tb{\tilde b}

\def\tst{\tilde t}
\def\ttau{\tilde \tau}

\def\tg{\tilde g}
\def\tnu{\tilde\nu}

\def\tq{\tilde q}
\def\tw{\widetilde W}
\def\tz{\widetilde Z}
\def\alt{\stackrel{<}{\sim}}
\def\agt{\stackrel{>}{\sim}}
\def\be{\begin{equation}}  
\def\ee{\end{equation}}  
\def\bea{\begin{eqnarray}}  
\def\eea{\end{eqnarray}}  
\def\beas{\begin{eqnarray*}}  
\def\eeas{\end{eqnarray*}}  
\newcommand\prd[3]{{\it Phys.\ Rev.\ }{\bf D #1} (#2) #3}

\newcommand\prl[3]{{\it Phys.\ Rev.\ Lett.\ }{\bf #1} (#2) #3}
\newcommand\plb[3]{{\it Phys.\ Lett.\ }{\bf B #1} (#2) #3}
\newcommand\jhep[3]{{\it J. High Energy Phys.\ }{\bf #1} (#2) #3}

\newcommand\npb[3]{{\it Nucl.\ Phys.\ }{\bf B #1} (#2) #3}
\newcommand\epjc[3]{{\it Eur.\ Phys.\ J. }{\bf C #1} (#2) #3}
\newcommand\ptp[3]{{\it Prog.\ Theor.\ Phys.\ }{\bf #1} (#2) #3}
\newcommand\zpc[3]{{\it Z.\ Physik }{\bf C #1} (#2) #3}
\newcommand{\hepph}[1]{hep-ph/#1}

\newcommand\ppnp[3]{{\it Prog.\ Part.\ Nucl.\ Phys.}{\bf  #1} (#2) #3}

\begin{document}
\begin{titlepage}
%\begin{flushright}
%OU-HEP/121030
%\end{flushright}

\vspace{0.5cm}
\begin{center}
{\Large \bf Electroweak versus high scale finetuning\\
in the 19-parameter SUGRA model
}\\ 
\vspace{1.2cm} \renewcommand{\thefootnote}{\fnsymbol{footnote}}
{\large Howard Baer$^1$\footnote[1]{Email: baer@nhn.ou.edu },
Vernon Barger$^2$\footnote[2]{Email: barger@pheno.wisc.edu },
and Maren Padeffke-Kirkland$^1$\footnote[3]{Email: m.padeffke@ou.edu } 
}\\ 
\vspace{1.2cm} \renewcommand{\thefootnote}{\arabic{footnote}}
{\it 
$^1$Dept. of Physics and Astronomy,
University of Oklahoma, Norman, OK 73019, USA \\
}
{\it 
$^2$Dept. of Physics,
University of Wisconsin, Madison, WI 53706, USA \\
}

\end{center}

\vspace{0.5cm}
\begin{abstract}
\noindent 
Recently, two measures of electroweak finetuning (EWFT) have been introduced for SUSY models:
$\Delta_{EW}$ compares the $Z$ mass to each separate weak scale contribution to $m_Z$ 
while $\Delta_{HS}$ compares the $Z$ mass to high scale input parameters and their 
consequent renormalizaton group evolution ($1/\Delta$ is the \% of fine tuning).
While the paradigm mSUGRA/CMSSM model has been shown to be highly finetuned under both 
parameters ($\Delta_{EW}\agt 10^2$ and $\Delta_{HS}\agt 10^3$),  the two-parameter non-universal Higgs model 
(NUHM2) in the context of radiatively-driven natural SUSY (RNS)
enjoys $\Delta_{EW}$ as low as 10, while $\Delta_{HS}$ remains $\agt 10^3$.
We investigate finetuning in the 19-free-parameter SUGRA model (SUGRA19). 
We find that with 19 free parameters, the lowest $\Delta_{EW}$ points are comparable to 
what can be achieved in NUHM2 with just 6 free parameters.
However, in SUGRA19, $\Delta_{HS}$ can now also reach as low as $5-10$. 
The conditions which lead to low $\Delta_{HS}$ include $m_{H_u}\sim m_Z$ at the high scale, with 
non-universal gaugino masses $M_{1,2}\gg M_3$ also at $m_{GUT}$. 
The low $\Delta_{HS}$ models are severely constrained by $b\to s\gamma$ branching fraction.
In both cases of low $\Delta_{EW}$ and $\Delta_{HS}$, the superpotential $\mu$ parameter 
should be $\sim 100-300$ GeV. While SUSY models with low EWFT may or may not be discoverable at LHC, 
the predicted light higgsinos must show up at a linear $e^+e^-$ collider with $\sqrt{s}\agt 2|\mu |$.

\vspace*{0.8cm}
%\noindent PACS numbers: 12.60.Jv,14.80.Va,14.80.Ly

\end{abstract}

\end{titlepage}

\section{Introduction}

Supersymmetric models of particle physics are renown for providing an elegant solution to the daunting 
gauge hierarchy problem.  They also receive indirect experimental support from 1. 
the measured strengths of weak scale gauge couplings, which allow for unification at a scale 
$m_{GUT}\simeq 2\times 10^{16}$ GeV within the Minimal Supersymmetric Standard Model (MSSM) and 2. from the
measured value of the top quark mass, which is sufficiently high as to radiatively drive electroweak 
symmetry breaking (REWSB)\cite{rewsb}.
Along with these plaudits, 3. the recent discovery by Atlas\cite{atlas_h} and CMS\cite{cms_h} of a 
Higgs-like boson with mass $m_h\simeq 125$ GeV confirms predictions from models of 
weak scale supersymmetry\cite{wss} where (in the context of the MSSM) a value $m_h\sim 114-135$ GeV 
was required\cite{mhiggs}. The emergent picture is that the MSSM (or possible extensions)
may provide a solid description of nature not only at the weak scale, but perhaps all the
way up to energy scales associated with grand unification\cite{primer}.

Such an audacious extrapolation has suffered a string of serious set-backs: 
so far, no signs of supersymmetric matter have emerged from LEP, LEP2, Tevatron or, more recently, LHC data. 
Recent analyses from Atlas and CMS in the context of the
minimal supergravity (mSUGRA or CMSSM) model\cite{msugra} 
require $m_{\tg}\agt 1.4$ TeV for $m_{\tq}\sim m_{\tg}$ and $m_{\tg}\agt 1$ TeV for $m_{\tg}\ll m_{\tq}$. 
Naively, these results exacerbate the so-called Little
Hierarchy Problem (LHP)\cite{LHP}: why is there an apparent discrepancy between the weak scale (typified by
$m_Z\simeq 91.2$ GeV) and the SUSY scale, where $m(sparticle)\agt 1$ TeV? The growing scale mismatch 
has led some physicists to call into question whether or not weak scale SUSY really exists, or at least to
concede that it suffers unattractive electroweak finetunings (EWFT)\cite{shifman}.

Traditionally, EWFT has been quantified using the Barbieri-Giudice 
measure\cite{ellis,bg,kane,ac,dg,ellis2,hb,king,fp,nilles,abe,martin,ant,perel,feng}
\be
\Delta_{BG}\equiv max_i\left|\frac{\partial\ln m_Z^2}{\partial\ln a_i}\right|
\ee 
where $a_i$ represents various fundamental parameters of the theory, usually taken to be some
set of soft SUSY breaking parameters defined at some high energy scale $\Lambda_{HS}$ below which the
theory in question is posited to be the correct effective field theory description of nature.
$1/\Delta$ is the \% of fine tuning.
The value of $\Delta_{BG}$ then answers the question: how stable is the fractional $Z$-boson mass
against fractional variation of high scale model parameters? 
Depending on which parameters are included in the set $a_i$, very different answers emerge\cite{feng}.
In addition, theories which are defined at very different values of $\Lambda_{HS}$, but which nonetheless
lead to exactly the same weak scale sparticle mass spectra, give rise to very different values of $\Delta_{BG}$.

\subsection{$\Delta_{EW}$ and $\Delta_{HS}$}

Recently, two different measures of EWFT-- $\Delta_{EW}$ and $\Delta_{HS}$-- 
have been proposed which answer a different but related question: 
how is it possible that $m_Z$ has a value of just 91.2 GeV while gluino and squark masses 
exist at TeV or even far beyond values? 
The answer should be: those independent contributions which enter the scalar potential and conspire to
build up the $Z$-boson mass should all be comparable to $m_Z$. 

Minimization of the scalar potential in the MSSM\cite{wss} leads to the well-known relation that
\be 
\frac{m_Z^2}{2} = \frac{m_{H_d}^2 + \Sigma_d^d - (m_{H_u}^2+\Sigma_u^u)\tan^2\beta}{\tan^2\beta -1} -\mu^2 \;,
\label{eq:mZs}
\ee 
where $m_{H_u}^2$ and $m_{H_d}^2$ are soft SUSY breaking (not physical) Higgs mass terms, 
$\mu$ is the superpotential Higgsino mass term, $\tan\beta\equiv v_u/v_d$ is the ratio of Higgs field vevs 
and $\Sigma_u^u$ and $\Sigma_d^d$ include a variety of independent radiative corrections\cite{rns}.

\subsubsection{$\Delta_{EW}$}

Noting that all entries in Eq.~\ref{eq:mZs} are defined at the weak scale, 
the {\rm electroweak fine-tuning parameter} 
\be 
\Delta_{EW} \equiv max_i \left|C_i\right|/(m_Z^2/2)\;, 
\ee 
may be constructed, where $C_{H_d}=m_{H_d}^2/(\tan^2\beta -1)$, $C_{H_u}=-m_{H_u}^2\tan^2\beta /(\tan^2\beta -1)$ and $C_\mu =-\mu^2$. 
Also, $C_{\Sigma_u^u(k)} =\Sigma_u^u(k)/(m_Z^2/2)$ and $C_{\Sigma_d^d(k)}=\Sigma_d^d(k)/(m_Z^2/2)$, 
where $k$ labels the various loop contributions included in Eq. \ref{eq:mZs}.
A low value of $\delew$ means less fine-tuning, {\it e.g.} $\Delta_{EW}=20$ corresponds to $\Delta_{EW}^{-1}=5\%$ finetuning
amongst terms contributing to $m_Z^2/2$.
Since $C_{H_d}$ and $C_{\Sigma_d^d(k)}$ terms are suppressed by $\tan^2\beta -1$, 
for even moderate $\tan\beta$ values the expression Eq. \ref{eq:mZs} reduces approximately to
\be 
\frac{m_Z^2}{2} \simeq -(m_{H_u}^2+\Sigma_u^u)-\mu^2\;.
\label{eq:approx}
\ee 
In order to achieve low $\delew$,  it is necessary that $-m_{H_u}^2$, $-\mu^2$ and each contribution to 
$-\Sigma_u^u$ all be nearby to $m_Z^2/2$ to within a factor of a few. 

A scan over mSUGRA/CMSSM parameter space, requiring that LHC sparticle mass constraints and $m_h=125\pm 2$ GeV 
be obeyed, finds a minimal value of $\Delta_{EW}\sim 10^2$, with more common values being $\Delta_{EW}\sim 10^3-10^4$.
Thus, one may conclude that the $Z$ mass is rather highly finetuned in this paradigm model. 
In the case of mSUGRA, the value $C_\mu$ becomes low only in the hyperbolic branch/focus point\cite{hb,fp} (HB/FP) region.
In this region, however, $m_0$ and consequently $m_{\tst_{1,2}}$ are very large, so that $\Sigma_u^u(\tst_{1,2})$ are 
each large, and the model remains finetuned.

Alternatively, if one moves to the two-parameter non-universal Higgs model (NUHM2)\cite{nuhm2}, with free parameters
\be
m_0,\ m_{1/2},\ A_0,\ \tan\beta ,\ \mu,\ m_A
\ee
then 
\begin{enumerate}
\item $\mu$ can be chosen in the $100-300$ GeV range since it is now a free input parameter, 
\item  a value of $m_{H_u}^2(m_{GUT})\sim (1.3-2.5)m_0$ may be chosen so that $m_{H_u}^2$ is driven only slightly
negative at the weak scale, leading to $m_{H_u}^2(weak)\sim -m_Z^2/2$, and 
\item with large stop mixing from $A_0\sim \pm 1.6 m_0$, the top-squark radiative corrections are softened
while $m_h$ is raised to the $\sim 125$ GeV level\cite{rns}.
\end{enumerate}
In the NUHM2 model, $\Delta_{EW}$ as low as $5-10$ can be generated. 
For such cases, the Little Hierarchy Problem seems to disappear. 
The low $\Delta_{EW}$ models are typified by the presence of light higgsinos 
$m_{\tw_1}^\pm,\ m_{\tz_{1,2}}\sim 100-300$ GeV which should be accessible to a linear $e^+e^-$ collider operating 
with $\sqrt{s}\agt 2|\mu |$. 
Also, $m_{\tg}\sim 1-5$ TeV while $m_{\tst_1}\sim 1-2$ TeV and $m_{\tst_2}\sim 2-4$ TeV.

The measure $\Delta_{EW}$ listed above is created from weak
scale MSSM parameters and so contains no information about any possible high scale origin, even though low
values of $\Delta_{EW}$ may be required of high scale models: in this sense, low $\Delta_{EW}$ captures a 
{\it minimal} EWFT required of even high scale SUSY models.

\subsubsection{$\Delta_{HS}$}

To include explicit dependence on the high scale $\Lambda$ at which the SUSY theory may be defined, 
we may write the {\it weak scale} parameters $m_{H_{u,d}}^2$ and $\mu^2$ in Eq.~(\ref{eq:mZs}) as 
\be
m_{H_{u,d}}^2= m_{H_{u,d}}^2(\Lambda) +\delta m_{H_{u,d}}^2; \ \ \ \
\mu^2=\mu^2(\Lambda)+\delta\mu^2\;,
\ee 
where
$m_{H_{u,d}}^2(\Lambda)$ and $\mu^2(\Lambda)$ are the corresponding
parameters renormalized at the high scale $\Lambda$. 
It is the $\delta m_{H_{u,d}}^2$ terms that will contain the $\log\Lambda$ dependence 
emphasized in constructs of natural SUSY models\cite{kn,papucci,brust}. 
In this way, we write
\be 
\frac{m_Z^2}{2} = \frac{(m_{H_d}^2(\Lambda)+ \delta m_{H_d}^2 +
\Sigma_d^d)-(m_{H_u}^2(\Lambda)+\delta m_{H_u}^2+\Sigma_u^u)\tan^2\beta}{\tan^2\beta -1} 
-(\mu^2(\Lambda)+\delta\mu^2)\;.
\label{eq:mZs_hs}
\ee 
In the same spirit used to construct $\Delta_{EW}$, 
we can now define a fine-tuning measure that encodes the
information about the high scale origin of the parameters by requiring
that each of the terms on the right-hand-side of Eq.~(\ref{eq:mZs_hs}) 
(normalized again to $m_Z^2/2$) be smaller than a value $\Delta_{\rm HS}$. 
The high scale fine-tuning measure $\Delta_{\rm HS}$ is thus defined to be
\be 
\Delta_{\rm HS}\equiv max_i |B_i |/(m_Z^2/2)\;, 
\label{eq:hsft} 
\ee 
with $B_{H_d}\equiv m_{H_d}^2(\Lambda)/(\tan^2\beta -1)$ etc., defined analogously to the set $C_{i}$. 

As discussed above, in models such as mSUGRA whose domain of validity extends to very high scales,
because of the large logarithms one would expect that (barring seemingly
accidental cancellations) the $B_{\delta H_u}$ contributions to $\Delta_{\rm HS}$ would be much larger than any
contributions to $\Delta_{\rm EW}$ because the term $m_{H_u}^2$ evolves from large $m_0^2$ through zero 
to negative values in order to radiatively break electroweak symmetry.
Thus, $\Delta_{HS}$ is numerically very similar to the EWFT measure advocated by Kitano-Nomura\cite{kn}
where $\Delta_{KN}=\delta m_{H_u}^2/(m_h^2/2)$

Scans of the mSUGRA/CMSSM model in Ref. \cite{sugra} found $\Delta_{HS}\agt 10^3$.
In Ref. \cite{rns}, scans over NUHM2 model similarly found $\Delta_{HS}\agt 10^3$.
Thus, both the mSUGRA and NUHM2 models would qualify as highly EW finetuned under $\Delta_{HS}$.

\subsection{Goals of this paper}

In this paper, we would like to maintain the SUSY grand desert scenario where the MSSM is postulated as the
correct effective theory below $Q\simeq m_{GUT}$. However, we would like to expand our set of input
parameters, in this case, to a maximal set of 19, which maintains the scenario of minimal flavor and minimal
$CP$-violation. The resulting model, dubbed here as SUGRA19\cite{bbs}, has the same parameter freedom as the
more popular pMSSM model\cite{pmssm}. However, unlike pMSSM defined at the weak scale, SUGRA19  maintains
the successes of renormalization group evolution, and its consequent gauge coupling unification and radiative
electroweak symmetry breaking due to the large value of $m_t$.

In this paper, we have several goals. 
The first is to check, under models with maximal parameter freedom, 
whether even lower values of $\Delta_{EW}\ll 10$ can be found, or whether NUHM2 already achieves the minimal
EWFT values. We will find that $\Delta_{EW}\sim 5-10$ is about as low as can be achieved while
maintaining accord with phenomenological constraints, and that the resulting models tend to look 
phenomenologically rather similar to RNS models as derived from NUHM2 parameter space. 

Our second goal is to check whether values of $\Delta_{HS}\ll 10^3$ can be found.  
In the case of SUGRA19, we will find that $\Delta_{HS}$ as low as $5-10$ can also be found, but only for 
special choices of non-universal SUGRA parameters. 
The low $\Delta_{HS}$ models appear tightly constrained if accord 
with $BF(b\to s\gamma )$ measurements is imposed.
Both the low $\Delta_{EW}$ and the low $\Delta_{HS}$ models are
characterized by light higgsinos with mass $\sim 100-300$ GeV, which should be accessible to linear
collider searches, although perhaps not accessible to LHC searches.
We present some interesting benchmark (BM) points for both low $\Delta_{EW}$ and low $\Delta_{HS}$.

Before proceeding, we mention our results in relation to several previous works on reduced EWFT in 
models with non-universal soft SUSY breaking terms. After several initial studies using the BG
measure in universal models\cite{ellis,bg,kane,ac,ellis2}, Kane and King\cite{king} showed that
non-universal gaugino masses with $M_1,\ M_2 >M_3$ lead to reduced EWFT. 
These were followed by similar studies by Lebedev {\it et al.}\cite{nilles}, Abe {\it et al.}\cite{abe}, Martin\cite{martin} 
(the latter of which was used to motivate a compressed SUSY mass spectrum) and Antusch {\it et al.}\cite{ant}. 
Recently, Gogoladze {\it et al.}\cite{shafi} studied the 
measures $\Delta_{EW}$ and $\Delta_{HS}$ within the context of GUT-motivated gaugino mass non-universality
using effectively a 5-parameter model with gaugino masses related by $M_1=\frac{2}{5}M_3+\frac{3}{5}M_2$.
In their study, they were already able to reduce the maximum $\Delta_{HS}$ values down to 
the $\sim 30$ level, which is already quite close to what is achieved here using 19 free parameters.

\section{Scan over 19 parameter SUGRA model}

To calculate superparticle mass spectra in the SUGRA19 model, we employ the
Isajet 7.83~\cite{isajet} SUSY spectrum generator Isasugra\cite{isasugra}. 
Isasugra begins the calculation of the sparticle mass spectrum with input $\overline{DR}$
gauge couplings and $f_b$, $f_\tau$ Yukawa couplings at the scale
$Q=M_Z$ ($f_t$ running begins at $Q=m_t$) and evolves the 6 couplings
up in energy to scale $Q=M_{\rm GUT}$ (defined as the value $Q$ where
$g_1=g_2$) using two-loop RGEs.  We do not enforce the exact
unification condition $g_3 = g_1 = g_2$ at $M_{\rm GUT}$, since a few
percent deviation from unification can be attributed to unknown
GUT-scale threshold corrections~\cite{Hisano:1992jj}.  Next, we impose
the SSB boundary conditions at $Q=M_{\rm GUT}$ and evolve the set
of 26 coupled two-loop MSSM RGEs~\cite{mv,yamada} back down in
scale to $Q=M_Z$.  Full two-loop MSSM RGEs are used for soft term
evolution, and the gauge and Yukawa coupling evolution includes
threshold effects in the one-loop beta-functions, so the gauge and
Yukawa couplings transition smoothly from the MSSM to SM effective
theories as different mass thresholds are passed.  In Isasugra, the
values of SSB terms which mix are frozen out at the scale 
$Q = m_{SUSY} =\sqrt{m_{\tst_L} m_{\tst_R}}$, while non-mixing SSB terms are
frozen out at their own mass scale~\cite{isasugra}.  The scalar
potential is minimized using the RG-improved one-loop MSSM effective
potential evaluated at an optimized scale $Q=\msusy$ to account
for leading two-loop effects~\cite{haber}.  Once the tree-level
sparticle mass spectrum is obtained,  one-loop radiative
corrections are calculated for all sparticle and Higgs boson masses,
including complete one-loop weak scale threshold corrections for the
top, bottom and tau masses at scale $Q=\msusy$~\cite{pbmz}.  Since
Yukawa couplings are modified by the threshold
corrections, the solution must be obtained iteratively, with
successive up-down running until a convergence at the required
level is found.  

We search for models with low $\Delta_{EW}$ and low $\Delta_{HS}$ 
by first performing a {\it broad-based} random scan over the following SUGRA19 parameter ranges:
\begin{itemize}
\item Gaugino masses: $M_1,\ M_2,\ M_3: 0-3.5$ TeV
\item First/second generation scalar masses: $m_{Q_1}$, $m_{U_1}$, $m_{D_1}$,
$m_{L_1}$, $m_{E_1}$: $0-3.5$ TeV,
\item Third generation scalar masses: $m_{Q_3}$, $m_{U_3}$, $m_{D_3}$,
$m_{L_3}$, $m_{E_3}$: $0-3.5$ TeV,
\item Higgs soft masses: $m_{H_u},\ m_{H_d}:\ 0-3.5$ TeV,
\item trilinear soft terms: $A_t,\ A_b,\ A_\tau$:$-3.5\ {\rm TeV}\ \to 3.5$ TeV,
\item ratio of weak scale Higgs vevs $\tan\beta :\ 2-60$.
\end{itemize}
We adopt a common mass for first and second generation scalars so as to avoid
the most stringent SUSY FCNC constraints\cite{masiero}. 

We require of our solutions that:
\bi
 \item electroweak symmetry be radiatively broken (REWSB),
 \item the neutralino $\tz_1$ is the lightest MSSM particle,
 \item the light chargino mass obeys the model independent LEP2 limit, 
$m_{\tw_1}>103.5$~GeV\cite{lep2ino} and
\item $123 <m_h< 128$~GeV.
\ei
We do not impose any LHC sparticle search limits since our general scan can
produce compressed spectra which in many cases can easily elude LHC gluino and squark searches.
Points which satisfy the above constraints are plotted as blue circles in the following scatter plots.

We will also calculate the values of $BF(b\to s\gamma )$\cite{vb_bsg,isabsg} and $BF(B_S\to\mu^+\mu^- )$\cite{isabmm} 
for each point generated. 
The measured value of $BF(b\to s\gamma )$ is found to be $(3.55\pm 0.26)\times 10^{-4}$~\cite{Asner:2010qj}.
For comparison, the SM prediction\cite{Misiak:2006zs} is $BF^{SM}(b\to s\gamma )=(3.15\pm 0.23)\times 10^{-4}$.
Also, recently the LHCb collaboration has found an excess over the
background for the decay $B_s\to\mu^+\mu^-$\cite{lhcb}.  They find a
branching fraction of $BF(B_s\to\mu^+\mu^- )=3.2^{+1.5}_{-1.2}\times
10^{-9}$ which is in accord with the SM prediction of $(3.2\pm 0.2)\times
10^{-9}$. 
Points with $BF(b\to s\gamma )$ within $3\sigma$ of its measured value $BF(b\to s\gamma )= (2.5-4.5)\times 10^{-4}$ 
and points with $BF(B_s\to\mu^+\mu^- )=(2-4.7)\times 10^{-9}$ will be labeled as light blue, showing
that these points are also in accord with $B$-physics constraints.

\begin{figure}[tbp]
\includegraphics[height=0.5\textheight]{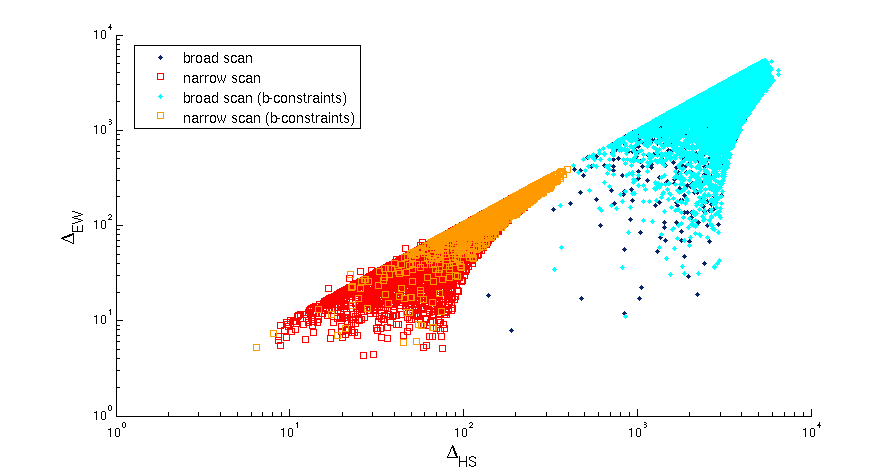}
\caption{Plot of $\Delta_{HS}$ vs. $\Delta_{EW}$ from a broad (dark/light blue) and 
focused (red/orange) scan over SUGRA19 model parameter space.
The orange and light blue points satisfy $B$-decay constraints while the dark blue and red points do not.
\label{fig:EW_HS}}
\end{figure}

Our first set of results are shown in Fig. \ref{fig:EW_HS}. 
The broad scan points are shown in blue. We see that the bulk of
generated points yield $\Delta_{EW}$ and $\Delta_{HS}\agt 10^3$, so would qualify 
as highly EW finetuned in generating $m_Z=91.2$ GeV. The points with the lowest
$\Delta_{EW}$ values come in with $\Delta_{EW}\sim 10$, which is similar to that which can be achieved
in the more restrictive NUHM2, but which is much better than what can be achieved in mSUGRA.

The lowest $\Delta_{EW}$ point has $\Delta_{EW}=7.9$, while the corresponding $\Delta_{HS}=190$.
The SUGRA19 parameters associated with this point are listed in Table \ref{tab:BMps} in the column
labeled as EW1. The point has the required low $\mu \sim 180$ GeV and 
$m_{H_u}^2(m_{weak})\sim -(171\ {\rm GeV})^2$. In addition, the large top-squark mixing 
$A_t/m_Q(3)\sim -2.1$ softens the top squark radiative corrections $\Sigma_u^u(\tst_{1,2})$ 
whilst raising $m_h$ up to $123.5$ GeV.

The corresponding sparticle mass spectra are listed in Table \ref{tab:mass}.
The gluinos and squarks are $\sim 2-3$ TeV: well beyond current LHC reach. The 
$\tw_2^\pm$ and $\tz_{1,2}$ are dominantly higgsino-like with a mass gap
$m_{\tz_2}-m_{\tz_1}\simeq 3$ GeV. Thus, even though the higgsinos can be produced with large
cross sections at LHC, the very soft visible energy release from their decays makes them difficult to
detect\cite{bbh}. The light higgsinos should be straightforward to detect at a linear
$e^+e^-$ collider with $\sqrt{s}\agt 400$ GeV. The lightest top squark $\tst_1$ has mass
less than 1 TeV: typically below values generated from radiative natural SUSY models\cite{rns}.
This leads to a somewhat anomalous branching fraction $BF(b\to s\gamma )\sim 2.5\times 10^{-4}$, 
below the measured value of $(3.55\pm 0.26)\times 10^{-4}$~\cite{Asner:2010qj}.
%
%\TABLE{
\begin{table}\centering
\begin{tabular}{lccc}
\hline
parameter & EW1 & HS1 & HS2 \\
\hline
$M_1(m_{GUT})$  & 2822.1 & 3266.2 & 3416.4  \\
$M_2(m_{GUT})$  & 3385.3 & 2917.8 & 3091.3  \\
$M_3(m_{GUT})$  & 884.9 & 1095.7 & 1085.8  \\
$m_Q(1)$      & 2484.7 & 1192.6 & 978.5  \\
$m_U(1)$      & 2506.2 & 2468.3 & 2440.6  \\
$m_D(1)$      & 2342.1 & 1508.9 & 1404.2  \\
$m_L(1)$      & 1820.4 & 623.8 & 754.8  \\
$m_E(1)$      & 1731.2 & 936.1 & 915.8  \\
$m_Q(3)$      & 698.3 & 6.6 & 371.3  \\
$m_U(3)$      & 1552.8 & 233.9 & 23.2  \\
$m_D(3)$      & 1498.5 & 2946.0 & 3052.4  \\
$m_L(3)$      & 3339.3 & 341.1 & 451.3  \\
$m_E(3)$      & 2114.9 & 1268.7 & 1247.5  \\
$m_{H_u}$      & 871.3 & 314.0 & 125.4 \\
$m_{H_d}$      & 2205.3 & 3160.4 & 2964.9  \\
$A_t$         & -1509.6 & -1024.4 & -801.3  \\
$A_b$         & 2301.7 & 3121.6 & 3294.3  \\
$A_\tau$       & 3307.3 & 1932.0 & 1754.5  \\
$\tan\beta$   & 27.0 & 51.1 & 29.0  \\
\hline
\hline
$\mu$      & 181.4 & 242.8 & 98.0  \\
$\Delta_{EW}$ & 7.9 & 17.9 & 5.2 \\
$\Delta_{HS}$ & 190.0 & 32.0 & 6.4 \\
\hline
\end{tabular}
\caption{Input parameters (GUT scale) in GeV for one low $\Delta_{EW}$ point and
two low $\Delta_{HS}$ points. We take $m_t=173.2$ GeV.
}
\label{tab:BMps}
\end{table}
%

%
%\TABLE{
\begin{table}\centering
\begin{tabular}{lccc}
\hline
mass (GeV) & EW1 & HS1 & HS2 \\
\hline
$m_{\tg}$   & 2042.9 & 2436.7 & 2428.8   \\
$m_{\tu_L}$ & 3650.7 & 2991.9 & 2968.5  \\
$m_{\tu_R}$ & 2980.5 & 3214.8 & 3191.6  \\
$m_{\te_R}$ & 2196.3 & 1763.6 & 1786.1  \\
$m_{\tst_1}$ & 879.5 & 1033.2 & 892.4   \\
$m_{\tst_2}$& 2305.1 & 1958.3 & 2394.9  \\
$m_{\tb_1}$ & 2121.8 & 1961.4 & 2418.0  \\
$m_{\tb_2}$ & 2327.7 & 2916.1 & 3495.8  \\
$m_{\ttau_1}$ & 2219.6 & 1049.5 & 1748.3  \\
$m_{\ttau_2}$ & 3865.8 & 1467.5 & 1911.3  \\
$m_{\tnu_{\tau}}$ & 3884.8 & 1464.9 & 1911.4  \\
$m_{\tw_2}$ & 2802.2  & 2393.0 & 2538.3  \\
$m_{\tw_1}$ & 192.1  & 255.5 & 104.1  \\
$m_{\tz_4}$ & 2810.2 & 2386.8 & 2530.3  \\ 
$m_{\tz_3}$ & 1261.2 & 1448.0 & 1513.5   \\ 
$m_{\tz_2}$ & 187.8 & 251.2 & 102.4  \\ 
$m_{\tz_1}$ & 184.7 & 247.9 & 99.3  \\ 
$m_A$      & 2759.7 & 2242.6 & 3176.4 \\
$m_h$       & 123.5 & 123.6 & 123.1  \\ 
\hline
$\Omega_{\tz_1}^{std}h^2$ & 0.007 & 0.013 & 0.003  \\
$BF(b\to s\gamma)\times 10^4$ & $2.5$  & $1.8$ & $2.6$  \\
$BF(B_s\to \mu^+\mu^-)\times 10^9$ & $3.9$  & $4.5$ & $3.8$  \\
$\sigma^{SI}(\tz_1 p)$ (pb) & $2.9\times 10^{-10}$  & $3.7\times 10^{-10}$ & $2.5\times 10^{-10}$ \\
%$\sigma^{SD}(\tz_1 p)$ (pb) & $2.1\times 10^{-4}$  & $2.9\times 10^{-5}$ \\
%$\langle\sigma v\rangle |_{v\to 0}$  (cm$^3$/sec) 
%& $2.8\times 10^{-25}$  & $3.1\times 10^{-25}$ \\
\hline
\end{tabular}
\caption{Sparticle masses in GeV and observables for one low $\Delta_{EW}$ and
two low $\Delta_{HS}$ points as in Table~\ref{tab:BMps}.
The measured values of the branching fractions are $BF(b\to s\gamma )=(3.55\pm 0.26)\times 10^{-4}$
and $BF(B_s\to\mu^+\mu^- )=3.2^{+1.5}_{-1.2}\times 10^{-9}$. 
}
\label{tab:mass}
\end{table}

\begin{figure}[tbp]
\includegraphics[height=0.3\textheight]{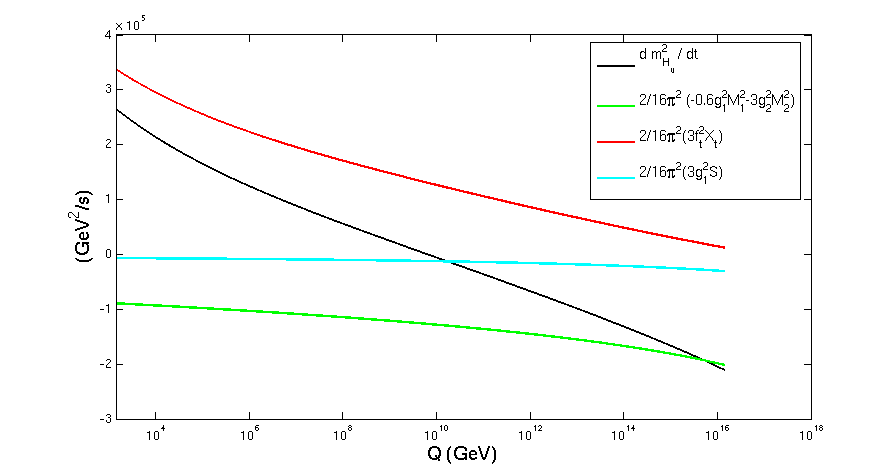}
\caption{Plot of {\it a}) slope $dm_{H_u}^2/dt$ vs. $Q$ 
from model HS1 with $\Delta_{HS}=32$.
\label{fig:slope}}
\end{figure}

A perhaps surprising result from Fig. \ref{fig:EW_HS} is that $\Delta_{HS}$ values far below
the NUHM2/mSUGRA minimal value of $10^3$ can now be found. In fact, the lowest 
$\Delta_{HS}$ point from the broad scan has a value of $ 32$, or 3.1\% EWFT, even including the effect of
high scale logarithms.
The parameter values for this point, labeled as HS1, are also listed in Table~\ref{tab:BMps}.
There are several features of the input parameters which lead to low $\Delta_{HS}$. 
First, the GUT scale value of $m_{H_u}^2=(314\ {\rm GeV})^2$, so our high scale starting point
for $m_{H_u}$ is not too far from $m_Z$. Second, the GUT scale gaugino masses $M_1$ and $M_2$
are $\sim 3 M_3\sim 3$ TeV. The RG running of $m_{H_u}^2$ is governed by
\be
\frac{dm_{H_u}^2}{dt}=\frac{2}{16\pi^2}\left(-\frac{3}{5}g_1^2M_1^2-3 g_2^2M_2^2+
\frac{3}{10}g_1^2 S +3 f_t^2X_t\right)
\ee
where $t=\log (Q^2/\mu^2)$, 
$S=m_{H_u}^2-m_{H_d}^2+Tr\left[{\bf m}_Q^2-{\bf m}_L^2-2{\bf m}_U^2+{\bf m}_D^2+{\bf m}_E^2\right]$ and
$X_t=m_{Q}^2(3)+m_U^2(3)+m_{H_u}^2+A_t^2$.
At $Q=m_{GUT}$, the large gaugino masses provide a large negative slope 
(green curve of Fig.~\ref{fig:slope}) for $m_{H_u}^2$, 
causing its value to increase while running towards lower mass scales. 
As the parameters evolve, $X_t$ increases due to the increasing squark soft terms so that the 
Yukawa coupling term grows (red curve from Fig. \ref{fig:slope}) and ultimately dominates; 
then $m_{H_u}^2$ is driven towards negative values, so that electroweak symmetry is finally broken. 
The total slope (black curve) passes through zero around $Q\sim 10^{10}$ GeV, indicating large cancellations in the
RG running of $m_{H_u}^2$.
Ultimately, the value
of $m_{H_u}^2 (m_{weak})\sim -(185\ {\rm GeV})^2$ so that both the starting and ending points
of $m_{H_u}^2$ remain not too far from $m_Z^2$, and hence $\delta m_{H_u}^2$ is not too far from $m_Z^2$, 
 thus fulfilling the most important condition required by low $\Delta_{HS}$.

\begin{figure}[tbp]
\includegraphics[height=0.3\textheight]{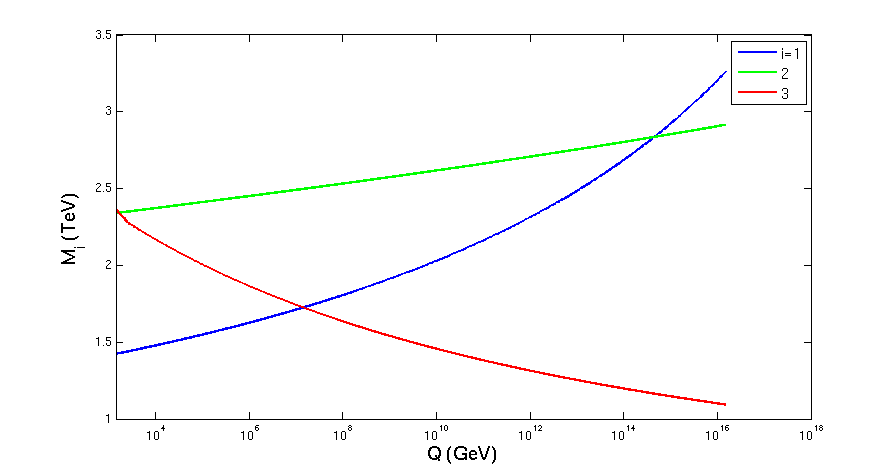}
\includegraphics[height=0.3\textheight]{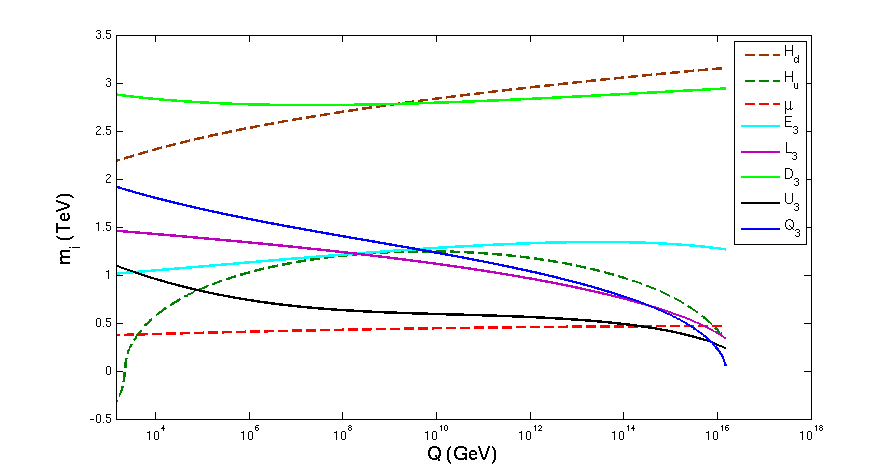}
\caption{Plot of {\it a}) running gaugino masses and 
{\it b}) running scalar masses vs. $Q$ from model HS1 with $\Delta_{HS}=32$.
\label{fig:run}}
\end{figure}

The RG running of gaugino masses and selected soft scalar masses for HS1 are shown in
Fig. \ref{fig:run}. In frame {\it a}), we see that indeed $M_1$ and $M_2$ start at $\sim 3$ TeV values and decrease,
whilst $M_3$ starts small at $Q= m_{GUT}$ and sharply increases. The gaugino mass boundary conditions then 
influence the running of the soft scalar masses in frame {\it b}). Most important is the running of $m_{H_u}^2$, 
which starts near $m_Z^2$ at $m_{GUT}$, runs up to about the TeV scale at $Q\sim 10^{10}$ GeV, and
then is pushed to small negative values by $Q\sim m_{weak}$. Also, $m_U(3)$ and $m_Q(3)$ start small, 
which aides the high $Q$ gaugino dominance in the running of $m_{H_u}^2$. By $Q\sim m_{weak}$, these
third generation squark soft terms have been pushed to the TeV scale. Thus, top squarks are not so heavy
and the radiative corrections $\Sigma_u^u(\tst_{1,2})$ are under control. 

\section{Results from narrow scan}

To hone in on SUGRA19 solutions with low $\Delta_{HS}$, we will impose a narrow, dedicated scan 
about our lowest $\Delta_{HS}$ solution:
\begin{itemize}
\item $M_1:\ 3-3.5$ TeV,\ $M_2:\ 2.7-3.2$ TeV,\ $M_3:\ 0.8-1.3$ TeV
\item $m_Q(1,2):\ 0.9-1.4$ TeV,\ $m_U(1,2):\ 2.2-2.7$ TeV,\ $m_D(1,2):\ 1.25-1.75$ TeV,\
$m_L(1,2):\ 0.4-0.9$ TeV,\ $m_E(1,2):\ 0.7-1.2$ TeV,
\item $m_Q(3):\ 0-0.5$ TeV,\ $m_U(3):\ 0-0.5$ TeV,\  $m_D(3):\ 2.7-3.2$ TeV,\  $m_L(3):\ 0.1-0.5$ TeV,\
$m_E(3):\ 1-2$ TeV,
\item $m_{H_u}:\ 0.05-0.55$ TeV,\  $m_{H_d}:\ 2.9-3.4$ TeV,
\item $A_t:\ -1.3\to -0.8$ TeV,\ $A_b:\ 2.9-3.4$ TeV,\ $A_{\tau}:\ 1.7-2.2$ TeV,
\end{itemize}
with $\tan\beta$ still $2-60$ as before.

\subsection{SUGRA19 parameters for low $\Delta_{HS}$ solutions}

The results from our narrow scan are shown in Fig. \ref{fig:EW_HS} as red $x$s,
while points that obey $B$-constraints are labeled as orange $x$s. 
The more focused sampling over lucrative parameter ranges has produced points with much 
lower $\Delta_{EW}$ values ranging down to $\sim 5$, and also solutions with $\Delta_{HS}$ as low as 6.
The $\Delta_{HS}=6.4$ solution is presented in Tables \ref{tab:BMps} and \ref{tab:mass} 
as BM model HS2. While point HS2 has $\mu$ of just 98 GeV, the lightest chargino
mass is $m_{\tw_1}=104.1$ GeV, slightly beyond the limit from LEP2 searches. 
Since gluino and squark masses are in the several TeV range, the point is also safe from 
LHC8 searches. The mass gaps $m_{\tw_1}-m_{\tz_1}=4.8$ GeV and $m_{\tz_2}-m_{\tz_1}=3.1$ GeV
so again there will be only tiny visible energy release from the higgsino decays.

\begin{figure}[tbp]
\includegraphics[height=0.5\textheight]{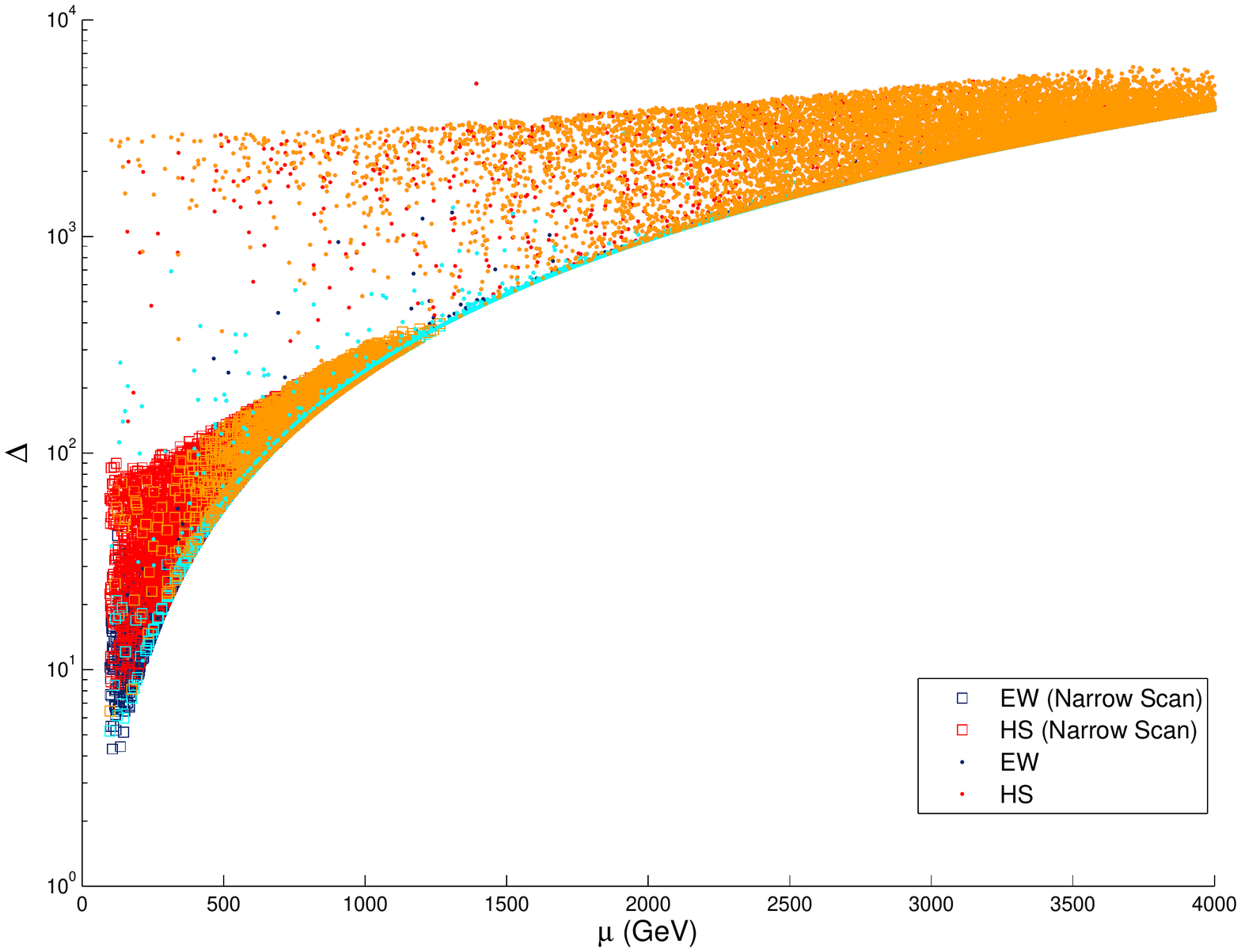}
\caption{Plot of $\Delta_{HS}$ and $\Delta_{EW}$  
vs. $\mu$ from scan over SUGRA19 model parameter space.
Color coding as in Fig. \ref{fig:EW_HS}.
\label{fig:mu}}
\end{figure}

To display the sort of parameter choices leading to low $\Delta_{HS}$, we show in
Fig. \ref{fig:mu} the values of $\Delta_{EW}$ (blue points) and $\Delta_{HS}$ (red/orange points) 
versus superpotential higgsino mass $\mu$ from the broad (circles) and narrow (x's)
scan. From the plot, we see unambiguously that low $|\mu |\sim m_Z$ is a necessary,
but not sufficient, condition to obtain {\it both} low $\Delta_{EW}$ and low $\Delta_{HS}$.
This translates into the solid prediction that four light higgsinos should lie
within reach of a linear $e^+e^-$ collider with $\sqrt{s}>2|\mu |$.

\begin{figure}[tbp]
\includegraphics[height=0.4\textheight]{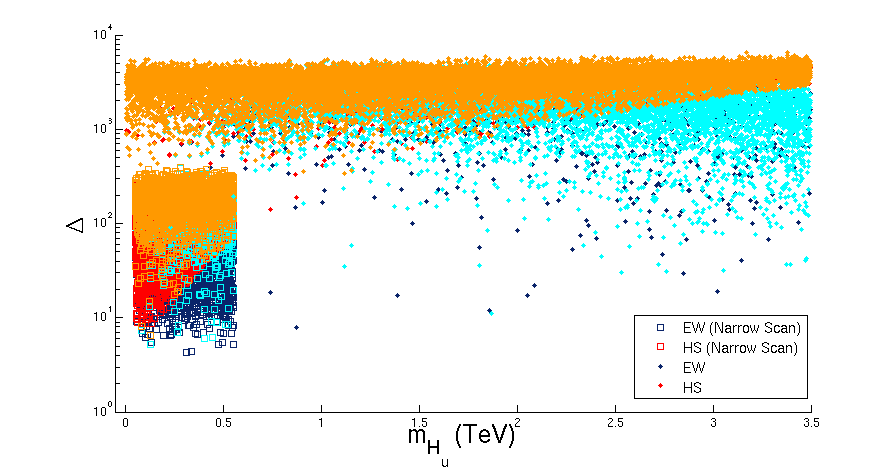}
\caption{Plot of $\Delta_{HS}$ and $\Delta_{EW}$  
vs. $m_{H_u} (m_{GUT})$ from scan over SUGRA19 model parameter space.
Color coding as in Fig. \ref{fig:EW_HS}.
\label{fig:mHu}}
\end{figure}

In Fig. \ref{fig:mHu}, we show $\Delta_{HS}$ and $\Delta_{EW}$  
vs. $m_{H_u} (m_{GUT})$ from the broad and narrow scans over SUGRA19 parameter space.
Here, we see that low $\Delta_{EW}$ solutions can be obtained over a large range of
$m_{H_u}(m_{GUT})$ values, as expected from radiative natural SUSY results\cite{rns}
which allow for a large cancellation between $m_{H_u}^2(m_{GUT})$ and $\delta m_{H_u}^2$.
However, the low $\Delta_{HS}$ solutions are only obtained for $m_{H_u}(m_{GUT})$ 
not too far from $m_Z$, as required by Eq. \ref{eq:mZs_hs}.

\begin{figure}[tbp]
\includegraphics[height=0.35\textheight]{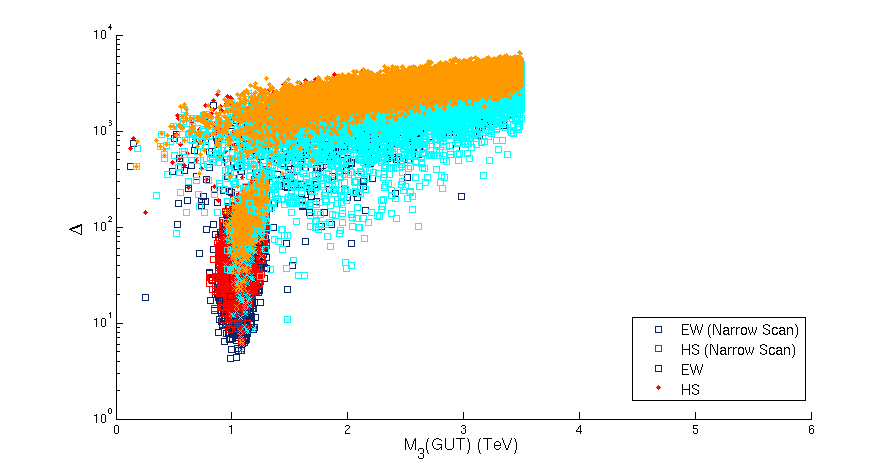}
\caption{Plot of $\Delta_{HS}$ and $\Delta_{EW}$  
vs. $M_3$ ($\sim m_{\tg}$) from scan over SUGRA19 model parameter space.
Color coding as in Fig. \ref{fig:EW_HS}.
\label{fig:M3}}
\end{figure}

In Fig. \ref{fig:M3}, we show $\Delta_{HS}$ and $\Delta_{EW}$  vs. $M_3$, where we note
that $m_{\tg}\simeq |M_3|$ up to radiative corrections. Low $\Delta_{EW}$ values allow
for $M_3\sim 1-3$ TeV, in accord with LHC searches which require $m_{\tg}\agt 1$ TeV for
not-too-compressed spectra.

\begin{figure}[tbp]
\includegraphics[height=0.3\textheight]{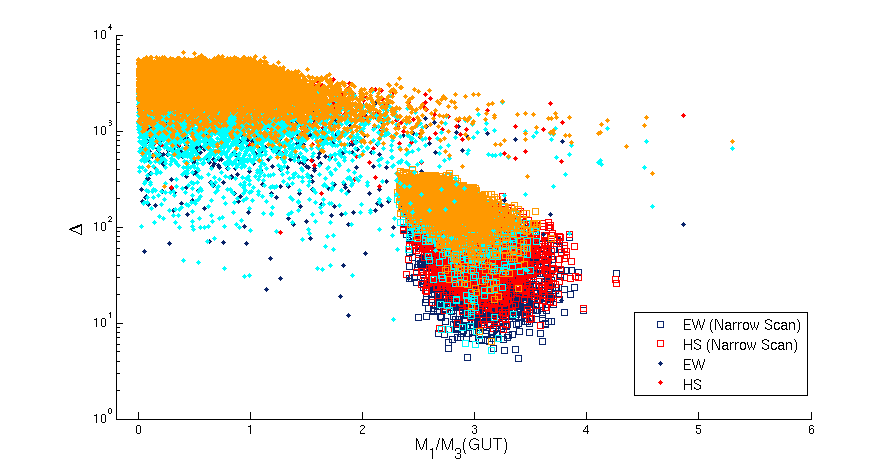}\\
\includegraphics[height=0.3\textheight]{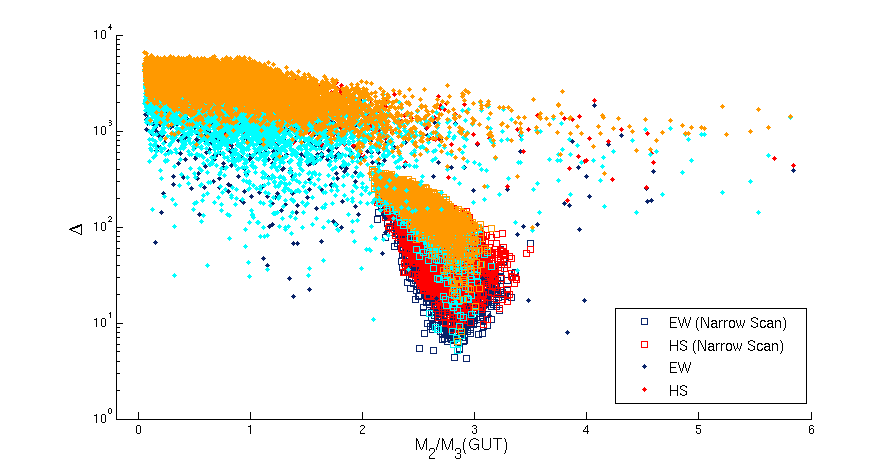}
\caption{Plot of $\Delta_{HS}$ and $\Delta_{EW}$  
vs. $M_1/M_3$ and $M_2/M_3$ from scan over SUGRA19 model parameter space.
Color coding as in Fig. \ref{fig:EW_HS}.
\label{fig:ino}}
\end{figure}

In the two frames of Fig. \ref{fig:ino}, we show  $\Delta_{HS}$ and $\Delta_{EW}$  
vs. {\it a}) $M_1/M_3$ and {\it b}) $M_2/M_3$, where all gaugino
masses are GUT scale values. The narrow scan has focused on the region around
$M_3\sim 1$ TeV, where solutions with $\Delta_{HS}\alt 10-100$ can be found.
From both frames, we see that non-universal GUT scale gaugino masses 
are required for low $\Delta_{HS}$ solutions, with ratios in the range
$M_1/M_3$ and $M_2/M_3\sim 2-4$ being preferred. As remarked earlier, the large
electroweak gaugino masses provide an initial upwards evolution of $m_{H_u}^2$
which is later cancelled by the downward push from the top Yukawa coupling at lower $Q$ values.

\subsection{Sparticle mass spectra from low $\Delta_{HS}$ solutions}

We have already seen from Fig. \ref{fig:mu} that very low values of $\mu\sim 100-300$ GeV 
are required for both low $\Delta_{HS}$ and low $\Delta_{EW}$ solutions. This translates
into the requirement of four light higgsino states $\tw_1^\pm$ and $\tz_{1,2}$ 
with mass $\sim 100-300$ GeV which are difficult to observe at LHC 
but should be observable at a linear $e^+e^-$ collider.

What of gluinos and squarks? We have already seen that $m_{\tg}\sim 1-3$ TeV is allowed for low 
$\Delta_{EW}$ solutions while low $\Delta_{HS}$ solutions prefer a lower value $m_{\tg}\sim 1$ TeV. 
This is due to the fact that too large a value of
$M_3(m_{GUT})$ (and hence a larger physical gluino mass) would cause an increase in 3rd generation squark masses,
which would increase the $X_t$ factor in the $m_{H_u}^2$ RGE, and cause a larger value of $\delta m_{H_u}^2$
to ensue.

\begin{figure}[tbp]
\includegraphics[height=0.35\textheight]{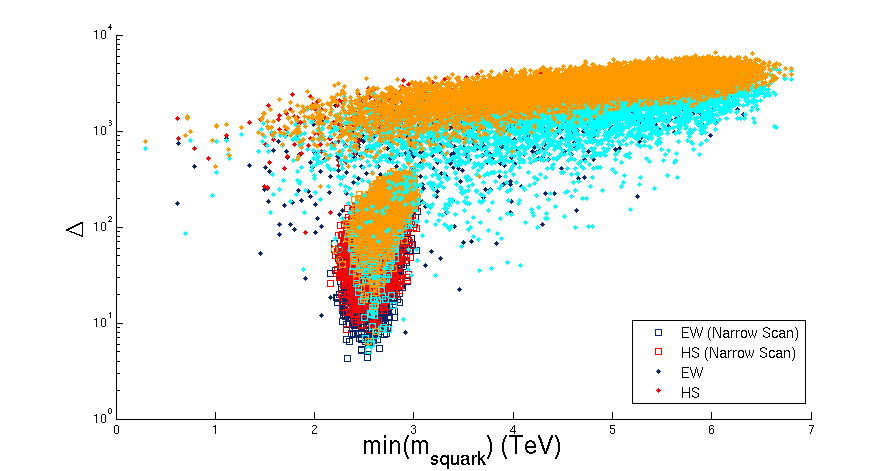}
\caption{Plot of $\Delta_{HS}$ and $\Delta_{EW}$  
vs. min($m_{\tq}$) from scans over SUGRA19 model parameter space.
Color coding as in Fig. \ref{fig:EW_HS}.
\label{fig:minq}}
\end{figure}

In Fig. \ref{fig:minq}, we show  $\Delta_{HS}$ and $\Delta_{EW}$
values versus min$[m_{\tq}]$, where the min is over 
all physical squark masses of the first two generations.
While low $\Delta_{EW}$ solutions allow for a broad range of $m_{\tq}\sim 1-5$ TeV, 
the low $\Delta_{HS}$ solutions tend to favor $m_{\tq}\sim 2-3$ TeV, beyond current LHC reach.

\begin{figure}[tbp]
\includegraphics[height=0.35\textheight]{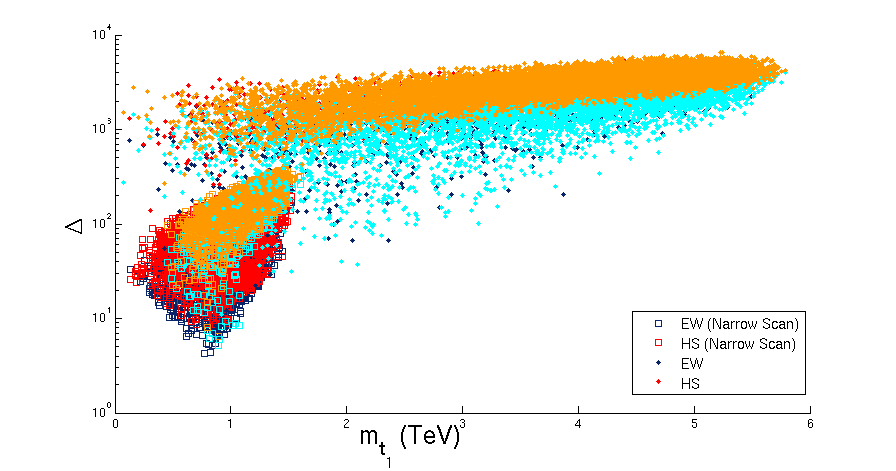}\\
\includegraphics[height=0.35\textheight]{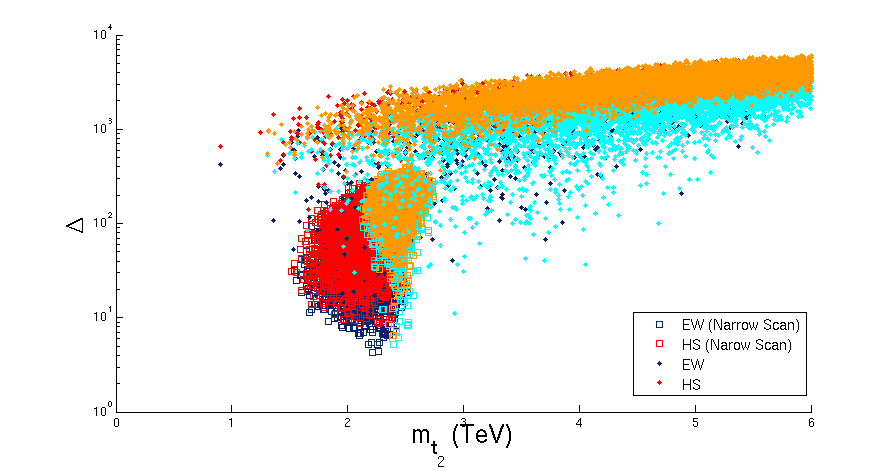}
\caption{Plot of $\Delta_{HS}$ and $\Delta_{EW}$  
vs. $m_{\tst_1}$ and $m_{\tst_2}$ from scans over SUGRA19 model parameter space.
Color coding as in Fig. \ref{fig:EW_HS}.
\label{fig:3rd}}
\end{figure}

In Fig. \ref{fig:3rd}, we show $\Delta_{HS}$ and $\Delta_{EW}$ versus 
top squark masses {\it a}) $m_{\tst_1}$ and {\it b}) $m_{\tst_2}$.
The lowest $\Delta_{HS}$ solutions are found for $m_{\tst_1}\sim 1$ TeV and $m_{\tst_2}\sim 2$ TeV.
These values are typically 1-2 TeV lower than results from RNS models
but still beyond most LHC reach projections for third generation squark detection.
The lowest $\Delta_{HS}$ solutions, colored dark blue and red, are usually in violation of the
$B$-constraints. This reflects the fact that low $\Delta_{HS}$ requires light higgsino-like charginos
(low $\mu$) and light stops, and hence there tend to be large anomalous contributions to the
$BF(b\to s\gamma )$ branching fraction. 

\begin{figure}[tbp]
\includegraphics[height=0.35\textheight]{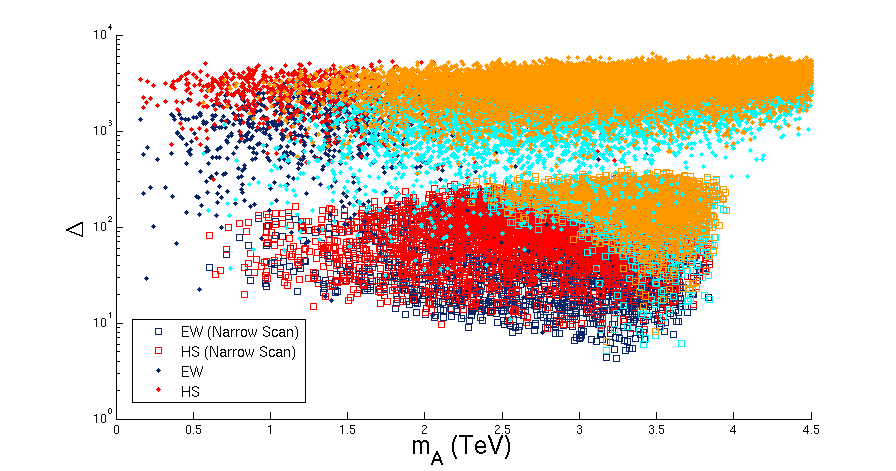}
\caption{Plot of $\Delta_{HS}$ and $\Delta_{EW}$  
vs. $m_A$ from scans over SUGRA19 model parameter space.
Color coding as in Fig. \ref{fig:EW_HS}.
\label{fig:mA}}
\end{figure}

In Fig. \ref{fig:mA}, we show $\Delta_{HS}$ and $\Delta_{EW}$ vs. $m_A$. Here, we see that 
the lowest $\Delta$ solutions favor $m_A$ (and hence $m_H$ and $m_{H^\pm}$) in the 2-4 TeV range, 
usually well beyond any projected LHC or ILC reach.

\subsection{Low $\Delta_{HS}$ solutions and $B$-physics constraints}

\begin{figure}[tbp]
\includegraphics[height=0.3\textheight]{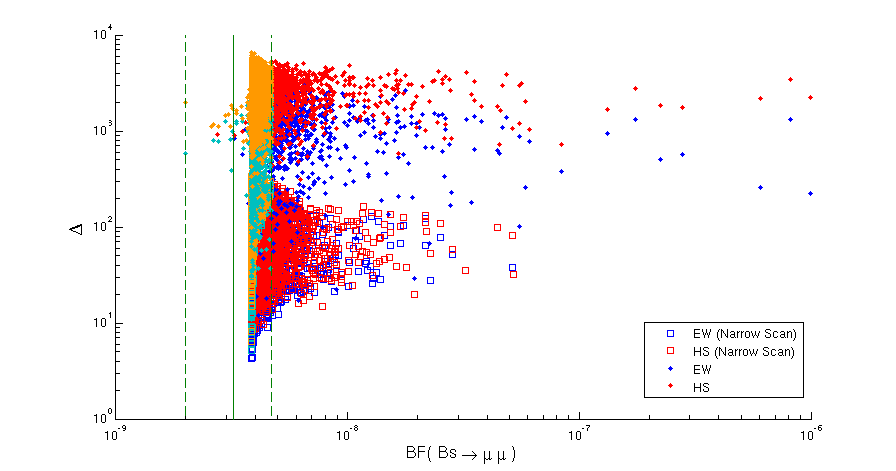}
\caption{Plot of $\Delta_{EW}$ and $\Delta_{HS}$
vs. $BF(B_s\to\mu^+\mu^- )$ from a 19 parameter scan.
Color coding as in Fig. \ref{fig:EW_HS}.
The vertical solid line is the measured value and the dashed lines are the 1$\sigma$ uncertainties.
\label{fig:bmm}}
\end{figure}

In Fig. \ref{fig:bmm}, we show values of $\Delta_{EW}$ and $\Delta_{HS}$
vs. $BF(B_s\to\mu^+\mu^- )$. 
The recent LHCb measurement\cite{lhcb} finds the
branching fraction of $BF(B_s\to\mu^+\mu^- )=3.2^{+1.5}_{-1.2}\times 10^{-9}$, 
in accord with the SM prediction of $(3.2\pm 0.2)\times 10^{-9}$. 
In supersymmetric models, this flavor-changing decay occurs
through pseudoscalar Higgs $A$ exchange\cite{isabmm}, and the
contribution to the branching fraction from SUSY is proportional to
$\frac{(\tan\beta)^6}{m_A^4}$.  
The decay is most constraining at large $\tan\beta$ and at low $m_A$. 
In the case of low $\Delta_{HS}$ solutions with lower $\tan\beta$ and heavier $m_A$, 
we find the bulk of solutions to lie within the newly measured error bars although
some solutions with large $\Delta_{HS}$ and $\Delta_{EW}$ will be excluded.

\begin{figure}[tbp]
\includegraphics[height=0.3\textheight]{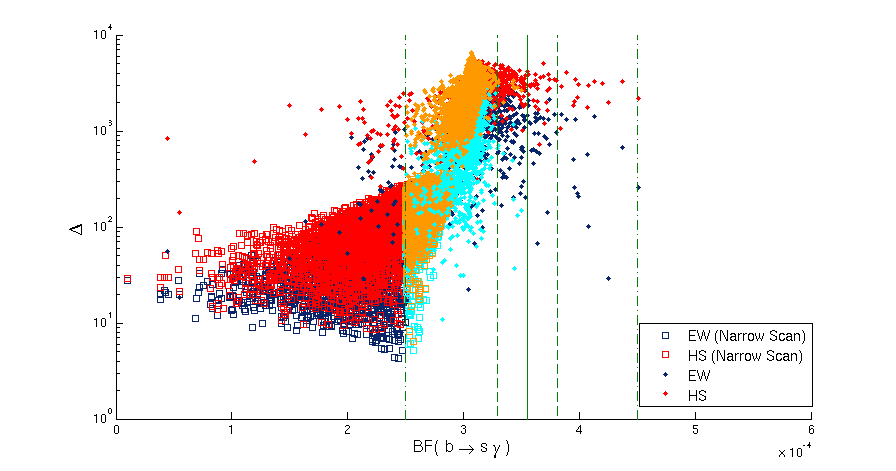}
\caption{Plot of $\Delta_{EW}$ and $\Delta_{HS}$
vs. $BF(b\to s\gamma )$ from a 19 parameter scan.
Color coding as in Fig. \ref{fig:EW_HS}.
The vertical solid line is the measured value and the dashed lines are the 1$\sigma$ and
3$\sigma$ uncertainties.
\label{fig:bsg}}
\end{figure}

In Fig. \ref{fig:bsg}, we show $\Delta_{EW}$ and $\Delta_{HS}$
vs. $BF(b\to s\gamma )$. We also show the measured central value and both 1 and 3-$\sigma$ error bars.
SUSY contributions to the
$b\to s\gamma$ decay rate come mainly from chargino-stop loops and
the W-charged Higgs loops, and so are large when these particles are light
and when $\tan\beta$ is large\cite{vb_bsg,isabsg}. 
In the case shown here, the low $\Delta_{HS}$ solutions which require third generation
squarks somewhat heavier than generic Natural SUSY but somewhat lighter than radiative Natural SUSY, 
we find the bulk of low $\Delta_{HS}$ solutions to lie a couple standard deviations below the
measured value. 

\subsection{Low $\Delta_{HS}$ solutions and dark matter}

\begin{figure}[tbp]
\includegraphics[height=0.3\textheight]{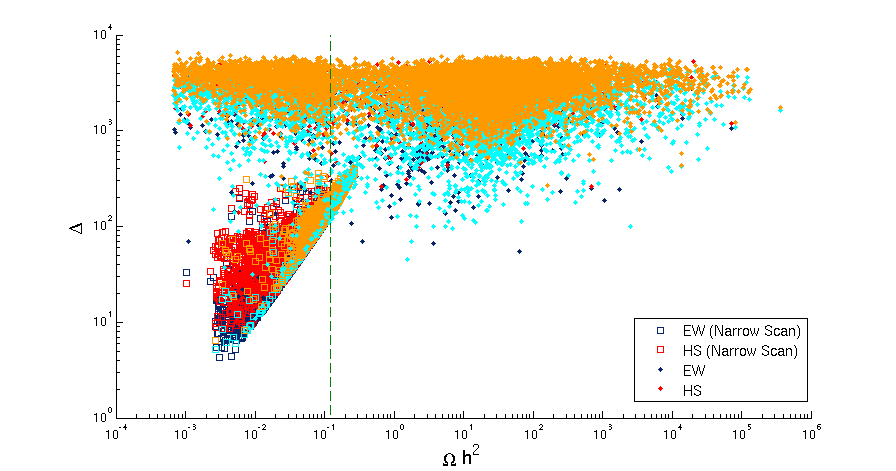}
\includegraphics[height=0.3\textheight]{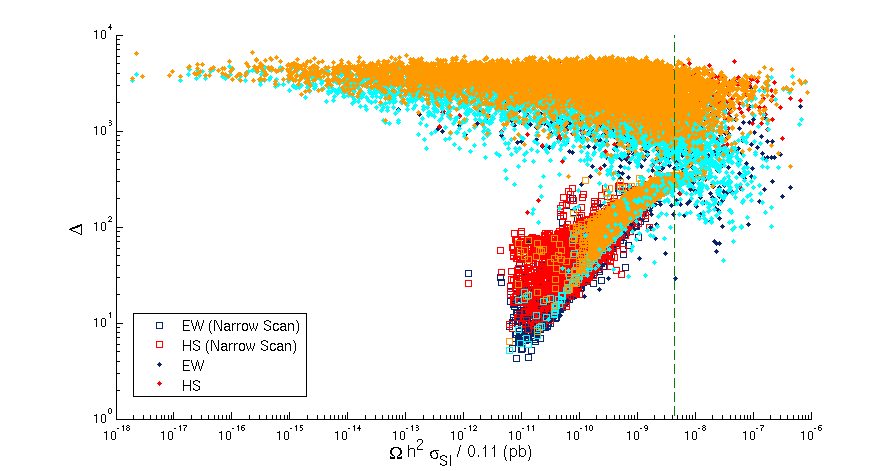}
\caption{Plot of $\Delta_{EW}$ and $\Delta_{HS}$
vs. {\it a}) standard neutralino relic abundance $\Omega_{\tz_1}^{std}h^2$ 
and {\it b}) rescaled spin-independent $\sigma (\tz_1 p)$ from a 19 parameter scan.
Color coding as in Fig. \ref{fig:EW_HS}.
The vertical dashed line in {\it a}) is the Planck measured value, while in {\it b}) 
it is the recent Xe-100 upper limit on spin-independent neutralino-proton scattering for a 150 GeV WIMP.
\label{fig:Oh2}}
\end{figure}

In this section, we show values of the relic neutralino abundance from our scan over
SUGRA19 parameter space. We use the IsaReD\cite{isared} relic density calculator from Isajet.
Our results are shown in Fig. \ref{fig:Oh2} frame {\it a}), where we plot
$\Delta_{HS}$ and $\Delta_{EW}$ vs. $\Omega_{\tz_1}h^2$. For comparison, we also show 
the location of the PLANCK-measured\cite{planck} relic density of dark matter (green dashed line).
Both the low $\Delta_{HS}$ and $\Delta_{EW}$ solutions populate a band located well below
the measured abundance. This reflects the fact that the low $\Delta_{HS,EW}$ solutions
all have low $\mu$ so that the $\tz_1$ is dominantly higgsino-like; these solutions enjoy an
ample annihilation cross section into $WW$, $ZZ$ etc. in the early universe.
Thus, the lowest $\Delta_{HS,EW}$ solutions are typically suppressed by factors of
$10-50$ below the measured dark matter abundance. Clearly, additional physics is needed
in the early universe to gain accord with experiment. One suggestion-- the presence 
of late-decaying scalar fields-- can either augment or diminish the relic 
abundance from its standard value\cite{gg}. Another possibility-- mixed neutralino
plus axion dark matter-- is favored by SUSY models with a standard underabundance of
neutralinos since thermal production of axinos and thermal/non-thermal
production of saxions followed by decays to SUSY particles can augment the 
neutralino abundance\cite{lessa} (depending on additional Peccei-Quinn parameters and the 
re-heat temperature $T_R$ after inflation). Any remaining gap in the generated neutralino abundance
relative to the measured abundance can be made up of axions.

In Fig. \ref{fig:Oh2}{\it b}), we plot the rescaled spin-independent neutralino-proton scattering
cross section $(\Omega_{\tz_1}^{std}h^2/0.11)\sigma^{SI}(\tz_1 p)$ from IsaReS\cite{isares}.
The prefactor accounts for the possibility  that the local dark matter abundance may be well below
the value assumed from neutralino-only CDM. For reference, we show the latest Xe-100 limit\cite{xe100}
(dashed green line) for $m_{\tz_1}=150$ GeV. The rescaled direct detection (DD) cross section 
is below current sensitivity by 1-2 orders of magnitude. 
This is lower than even DD projections from radiative natural SUSY\cite{bbm} since in that case the 
$\tz_1$ is mainly higgsino but with a non-negligible gaugino component. Since
the $\tz_1-\tz_1-h$ coupling depends on a product of gaugino-higgsino components, the value of 
$\sigma^{SI}(\tz_1 p)$ in the RNS case never gets too small.
In our current case, with non-universal gaugino masses, the $\tz_1$ can be a much more pure higgsino state, 
and consequently can have a significantly lower DD rate.

\section{Conclusions:} 

In previous studies, the radiative natural SUSY model has emerged as a way to reconcile low EWFT 
with lack of SUSY signals at LHC8 and the presence of a light Higgs scalar with mass $m_h\sim 125$ GeV.
The RNS model cannot be realized within the restrictive mSUGRA/CMSSM framework, but can be realized within the 
context of NUHM2 models (which depend on 6 input parameters) and where $\mu$ can be a free input value.
In RNS models, $\Delta_{EW}$ as low as $\sim 10$ can be generated while $\Delta_{HS}$ as low as $10^3$
can be found.

In this study, we have implemented scans over the most general minimal flavor- and minimal $CP$-violating
GUT scale SUSY model-- SUGRA19-- with two goals in mind. 
Our first goal was to check if the additional freedom of 13 extra parameters allows for much lower
$\Delta_{EW}$ solutions. 
In previous work-- by proceeding from mSUGRA to NUHM2 models-- 
a reduction in the minimum of $\Delta_{EW}$ of at least a factor of 10 was found\cite{sugra,rns}. 
In the present work, we do not find any substantial reduction in 
the minimal $\Delta_{EW}$ value by proceeding from the NUHM2 model to SUGRA19. 
The parameter freedom of NUHM2 appears sufficient to minimize $\Delta_{EW}$ to its lowest values of $\sim 5-10$.

Our second goal was to check whether the additional parameter freedom can improve on the
high scale EWFT parameter $\Delta_{HS}$. In this regard, we find improvements by factors ranging up to $\sim 150$!
In order to generate low values of $\Delta_{HS}$, one must generate $\mu\sim 100-300$ GeV as usual, 
but also one must start with $m_{H_u}^2\sim m_Z^2$ at the GUT scale, and then generate relatively
little change $\delta m_{H_u}^2$ during evolution from $m_{GUT}$ to $m_{weak}$. Small values of 
$\delta m_{H_u}^2$ can be found if one begins with electroweak gaugino masses $M_{1,2}\sim 3 M_3$ at the
GUT scale so that gaugino-induced RG evolution dominates at high $Q\sim m_{GUT}$. Then at lower
$Q$ values approaching the weak scale,  top-Yukawa terms dominate the running of $m_{H_u}^2$, leading to broken
electroweak symmetry, but also to not much net change in $m_{H_u}^2$ during its evolution from $m_{GUT}$ to $m_{weak}$.

The solutions with low $\Delta_{HS}$ are characterized by the presence of four light higgsinos
$\tw_1^\pm$ and $\tz_{1,2}$ similar to RNS models. However, in contrast to RNS models, the 
third generation squarks tend to be lighter (although not as light as generic natural SUSY which favors
$m_{\tst_{1,2}}\alt 500$ GeV). The lighter third generation squarks lead to significant SUSY contributions to
the decay $b\to s\gamma$, and seem to be disfavored by the measured value of this branching fraction.
In the case of low $\Delta_{HS}$ models, the lightest neutralino is more higgsino-like than
in RNS models, leading to even lower values of predicted relic density and low direct detection rates.
The remaining CDM abundance may be augmented by scalar field or axino/saxion production and decay in 
the early universe, and in the latter case, the additional presence of axions is expected.

\section*{Acknowledgments}

We thank P. Huang, D. Mickelson, A. Mustafayev and X. Tata for previous collaborations leading up to
this study.
This work was supported in part by the US Department of Energy, Office of High
Energy Physics.

%
%%%%%%%%%%%%%%%%%%%%%%%%%%%%%%%%%%%%%%%%%%%%%%%%%%%%%%

%
\end{document}